\documentclass[twocolumn,prd,nofootinbib,aps,prl,floats,floatfix,amsmath,amssymb,longbibliography,secnumarabic]{revtex4-1} %
\usepackage[english]{babel}
\usepackage{csquotes}
\usepackage{csquotes}

\usepackage[letterpaper,top=2cm,bottom=2cm,left=3cm,right=3cm,marginparwidth=1.75cm]{geometry}

\usepackage{amsmath}
\usepackage{graphicx,slashed}
\usepackage[colorlinks=true,citecolor=red]{hyperref}

\newcommand{\be}{\begin{equation}}
\newcommand{\ee}{\end{equation}}

\newcommand{\bea}{\begin{eqnarray}}
\newcommand{\eea}{\end{eqnarray}}
\newcommand{\nn}{\nonumber}

\begin{document}

\title{Comment on ``Spontaneous baryosynthesis with large initial
phase''}
\author{James M.\ Cline}
\email{jcline@physics.mcgill.ca}
\affiliation{McGill University Department of Physics \& Trottier Space Institute, 3600 Rue University, Montr\'eal, QC, H3A 2T8, Canada}

\begin{abstract}
Recently \url{https://arxiv.org/pdf/2512.11011} set out to improve on previous work from 1994 by Dolgov and Freese, who used a small-angle approximation to derive the yield of spontaneous baryogenesis from a rolling phase, a pseudo-Nambu-Goldstone boson 
coupled to the baryon current.  The goal of the recent paper was to investigate what happens when the small-angle approximation is not imposed.  I point out a serious technical shortcoming in their derivation.

\end{abstract}

\maketitle

Ref.\ \cite{Krasnov:2025fej} has re-examined a seminal model of spontaneous baryogenesis first proposed in
1994 by Ref. \cite{Dolgov:1994zq}.  The Lagrangian is
\bea
  {\cal L} &=&  \partial_\mu\Phi^*\partial^\mu\Phi -V(|\Phi|) + i\bar Q\slashed{\partial}Q + i\bar L\slashed{\partial}L\nn\\
  &+& g(\Phi \bar Q L + \Phi^*\bar L Q) + U(\theta)\,,
\eea
where $Q$ and $L$ are fermions carrying baryon and lepton number, respectively.  The scalar $\Phi$ gets a vacuum expectation value and can be approximated as
$\Phi\cong f/\sqrt{2}e^{i\phi/f}$.  Its phase
$\theta=\phi/f$
is a pseudo-Nambu-Goldstone boson (PNGB), whose mass is determined by a small $U(1)$ symmetry breaking term $U(\theta)$ in the potential.  

By a field redefinition
$Q\to e^{i\theta}Q$, the appearance of $\theta$ in the interaction Lagrangian can be moved out of the $\bar QL$ mass term and into the $Q$ kinetic term, generating the interaction $i\bar Q(\slashed{\partial}\theta) Q$.
Cosmological rolling of $\theta$ in its potential generates a chemical potential $\dot\theta$ for baryon number and thereby gives rise to spontaneous baryogenesis.

One deficiency with this argument, overlooked by Dolgov and Freese and by the current authors, is that $B-L$ is conserved by the Lagrangian, except for the term $U(\theta)$ that drives the asymmetry production.  Therefore an equal asymmetry in $L$ will be produced, leaving $B-L = 0$.  The rolling of $\theta$ should
therefore begin at low temperature, below the freezeout of electroweak sphalerons, to generate a lasting asymmetry.
Otherwise the sphalerons, which violate $B+L$, will drive both
asymmetries to zero.

The goal of Ref.\ \cite{Krasnov:2025fej} is to solve for the dynamics of $\theta$ and the subsequent baryon production, in the case where initially $\theta \sim \pi$ rather than starting at a small angle.  The small-angle approximation was made to simplify the derivation of the original estimates made in in Refs.\ \cite{Dolgov:1994zq,Dolgov:1996qq}.

The main technical problem with Ref.\ \cite{Krasnov:2025fej} appears in Section 
4, where they attempt to solve the equation of motion (EOM)
derived in Ref.\ \cite{Dolgov:1994zq} for homogeneous field configurations $\theta(t)$:
\be
    \ddot\theta + U'(\theta)/f^2 = -{1\over f^2}\langle \partial_\mu J_B^\mu \rangle = i{g\over f}\langle(\bar Q L-\bar L Q)\rangle\,.
    \label{EOM}
\ee
Here $J^\mu_B = \bar Q\gamma^\mu Q$ is the baryon current, and a mean-field approximation is used, calculating 
the expectation value in terms of the  $Q,\,L$ propagators in the $\theta(t)$ background.  This term accounts for the back-reaction of baryon production on the evolution of $\theta$, in a more exact way than adding a phenomenological $\Gamma\dot\theta$ damping term to the equation of motion.

After a lengthy derivation, the seminal Ref.\ \cite{Dolgov:1994zq} obtains  for the right-hand-side of Eq.\ (\ref{EOM}) the first and second lines of
\onecolumngrid
\bea
&&-4{g^2\over\pi^2}\lim_{\omega\to\infty}\int_0^\infty
 d\omega\,\omega^2\int_{-\infty}^0 dt'\,\sin(2\omega t')\,\sin\Delta\theta \nn\\
 \label{damping}
 &=&  - {g^2\over 2\pi^2}\lim_{\omega\to\infty}
 \int_{-\infty}^0 dt'\left[\cos(2\omega t')-1\over t'\right]\left(\ddot\theta(t+t')\cos\Delta\theta-\dot\theta^2(t+t')\sin\Delta\theta\right)\\
 &=& {g^2\over \pi^2} \lim_{\omega\to\infty}
 \int_{-\infty}^0 dt'\left[\sin^2\omega t'\over t'\right]
 {d\over dt'}\left(\dot\theta(t+t')\cos\Delta\theta\right)\,,\nn 
\eea
\twocolumngrid
\noindent where $\Delta\theta = \theta(t+t')-\theta(t)$.  Hence it becomes an integro-differential equation that would be difficult to solve exactly.  However, in the small-angle approximation, Dolgov and Freese were able to reduce it to a simple damping term $\Gamma\dot\theta$, where
$\Gamma = g^2 m_R/(4\pi)$, and $m_R$ is the renormalized mass of the PNGB.  Mass renormalization is required, since in this approximation, Eq.\ (\ref{damping}) also contributes a term to the bare mass squared from the $U(\theta)$ potential, that diverges as $\omega\to\infty$. 

The last line of Eq.\ (\ref{damping}) was derived by Ref.\ \cite{Krasnov:2025fej}, who proceeded to integrate it by parts.
Assuming the surface terms vanish, they then take
\bea
\lim_{\omega\to\infty}{d\over dt}\!\!\left[\sin^2\omega t'\over t'\right]\!\! &=& \!\!\lim_{\omega\to\infty}\!\! \left({\omega\sin 2\omega t'\over t'} - {\sin^2\omega t'\over t'^2}\right)\nn\\
&\stackrel{?}{=}& \lim_{\omega\to\infty} \pi\,\omega\, \delta(t')\,,
\label{weird}
\eea
using the fact that the two terms in the top line are representations of $2\pi\omega\delta(t')$ and $\pi\omega\delta(t')$ respectively.
I have added the question mark in the bottom relation, since it is clearly ill-defined.  By the same logic, one could alternatively write
\be
\lim_{\omega\to\infty} \pi\omega {d\over dt'}[t'\delta(t')]\,,
\ee
which leads to a different (equally ill-defined) result after integration by parts, namely zero.

In order to overcome the nonexistence of the limit in Eq.\ (\ref{weird}), the authors of Ref.\ \cite{Krasnov:2025fej}
propose that $\omega$ should take some large but finite value, whose physical meaning is never made clear.  This enables them to do the $t'$ integral trivially, and to reduce the ODE to a simple form that is local in time,
\be
    \ddot\theta + \Gamma\dot\theta + U'(\theta) = 0\,,
    \label{nonlinear}
\ee
where now the damping rate $\Gamma$ is linear in the unphysical
cutoff $\omega$, which was supposed to be taken to infinity.
Since they did not make the small angle approximation, $U(\theta)$ can be the full potential, typically of the form 
$\Lambda^4(1-\cos\theta)$ for a PNGB.  It is easy to solve Eq.\ (\ref{nonlinear}) numerically for different values of $\Gamma$ and initial angles $\theta_0$, which no longer needs to be small.
This is the content of Ref.\ \cite{Krasnov:2025fej}, which concludes that there is no dramatic dependence upon initial conditions of the baryon yield from spontaneous baryogenesis.  

While these modest results may be qualitatively correct when $\Gamma$ takes the Dolgov-Freese value $g^2 m_R/(4\pi)$, the means by which Ref.\ \cite{Krasnov:2025fej} obtained them are dubious, to 
put it politely.

\bigskip
I thank Ufuk Aydemir for useful discussions.

\bibliographystyle{utphys}
\bibliography{sample}

@article{Krasnov:2025fej,
    author = "Krasnov, Maxim and Khlopov, Maxim and Aydemir, Ufuk",
    title = "{Spontaneous baryosynthesis with large initial phase}",
    eprint = "2512.11011",
    archivePrefix = "arXiv",
    primaryClass = "hep-ph",
    month = "12",
    year = "2025"
}

@article{Dolgov:1994zq,
    author = "Dolgov, Alexandre and Freese, Katherine",
    title = "{Calculation of particle production by Nambu Goldstone bosons with application to inflation reheating and baryogenesis}",
    eprint = "hep-ph/9410346",
    archivePrefix = "arXiv",
    reportNumber = "UM-AC-94-35",
    doi = "10.1103/PhysRevD.51.2693",
    journal = "Phys. Rev. D",
    volume = "51",
    pages = "2693--2702",
    year = "1995"
}

@article{Dolgov:1996qq,
    author = "Dolgov, Alexandre and Freese, Katherine and Rangarajan, Raghavan and Srednicki, Mark",
    title = "{Baryogenesis during reheating in natural inflation and comments on spontaneous baryogenesis}",
    eprint = "hep-ph/9610405",
    archivePrefix = "arXiv",
    reportNumber = "ACT-09-96, TAC-96-015, UCSBTH-96-19, UM-AC-96-05",
    doi = "10.1103/PhysRevD.56.6155",
    journal = "Phys. Rev. D",
    volume = "56",
    pages = "6155--6165",
    year = "1997"
}

\end{document}